\documentclass[twocolumn,showpacs,preprintnumbers,amsmath,amssymb,prb,rapid]{revtex4}

\usepackage{epsfig}% Include figure files
\usepackage{graphicx}% Include figure files
\usepackage{dcolumn}% Align table columns on decimal point
\usepackage{bm}% bold math

\begin{document}

\title{Microscopic model for the magnetization plateaus in NH$_4$CuCl$_3$}

\author{Masashige Matsumoto}
\affiliation{Theoretische Physik, ETH-H\"onggerberg, CH-8093 Z\"urich, Switzerland}
\affiliation{Department of Physics, Shizuoka University, 836 Oya, Shizuoka 422-8529, Japan}

%\date{April 24, 2003}
\date{September 12, 2003}

\begin{abstract}
A simple model consisting of three distinct dimer sublattices is proposed
to describe the magnetism of NH$_4$CuCl$_3$.
It explains the occurrence of magnetization plateaus only at 1/4 and 3/4
of the saturation magnetization.
The field dependence of the excitation modes observed by ESR measurements
is also explained by the model.
The model predicts that the magnetization plateaus should disappear under high pressure.
\end{abstract}


\maketitle

%%%%%%%%%%%%%%%%%%%%%%%%%%%%%%%%%%%%%%%%%%%%%%%%%%%%%%%%%%%%%%%%%%%%%%%%%%%%%%%%
%Macros
%%%%%%%%%%%%%%%%%%%%%%%%%%%%%%%%%%%%%%%%%%%%%%%%%%%%%%%%%%%%%%%%%%%%%%%%%%%%%%%%
\newcommand{\K}{{KCuCl$_3$}}
\newcommand{\Tl}{{TlCuCl$_3$}}
\newcommand{\N}{{NH$_4$CuCl$_3$}}

\newcommand{\br}{{\mbox{\boldmath$r$}}}
\newcommand{\bA}{{\mbox{\boldmath$A$}}}
\newcommand{\bB}{{\mbox{\boldmath$B$}}}
\newcommand{\bH}{{\mbox{\boldmath$H$}}}
\newcommand{\bD}{{\mbox{\boldmath$D$}}}
\newcommand{\bk}{{\mbox{\boldmath$k$}}}
\newcommand{\bn}{{\mbox{\boldmath$n$}}}
\newcommand{\sk}{{\mbox{\footnotesize $k$}}}
\newcommand{\bsk}{{\mbox{\footnotesize \boldmath$k$}}}
\newcommand{\bsQ}{{\mbox{\footnotesize \boldmath$Q$}}}
\newcommand{\bsr}{{\mbox{\footnotesize \boldmath$r$}}}
\newcommand{\bS}{{\mbox{\boldmath$S$}}}
\newcommand{\bQ}{{\mbox{\boldmath$Q$}}}
\newcommand{\bd}{{\mbox{\boldmath$d$}}}
\newcommand{\bsd}{{\mbox{\footnotesize{\boldmath$d$}}}}
\newcommand{\bsigma}{{\mbox{\boldmath$\sigma$}}}
\newcommand{\ha}{{\hat{a}}}
\newcommand{\hb}{{\hat{b}}}
\newcommand{\hc}{{\hat{c}}}
\newcommand{\ex}{\eta_x}
\newcommand{\ey}{\eta_y}
\newcommand{\del}{\partial}
\newcommand{\tK}{\tilde{K}}

%%%%%%%%%%%%%%%%%%%%%%%%%%%%%%%%%%%%%%%%%%%%%%%%%%%%%%%%%%%%%%%%%%%%%%%%%%%%%%%%
%\section{introduction}
%%%%%%%%%%%%%%%%%%%%%%%%%%%%%%%%%%%%%%%%%%%%%%%%%%%%%%%%%%%%%%%%%%%%%%%%%%%%%%%%
\N, \Tl, and \K~are isostructual quantum spin systems
consisting of two-leg ladders separated by NH$_4^+$, Tl$^+$, and K$^+$ ions.
The two-leg ladders are composed of Cu$^{2+}$ ions with spin $S=1/2$
which interact antiferromagnetically (AF) through the Cl$^-$ ions.
These compounds can be considered as coupled two-leg
$S=1/2$ Heisenberg AF spin ladders,
and show various types of magnetization curves.
Shiramura {\it et al.} found that
\N~has magnetization plateaus at 1/4 and 3/4 of the saturation moment,
\cite{Shiramura-1998}
while \Tl~and \K~have no plateaus in their magnetization curves.
\cite{Shiramura-1997}
For \N, Kurniawan {\it et al.} performed ESR experiments
which revealed that \N~has a finite excitation gap in the plateau regions.
\cite{Kurniawan-1}
By studying specific heat, they found AF order already at zero field
and suggested a phase diagram in finite fields
containing three magnetically ordered phases.
\cite{Kurniawan-2}

The origin of magnetization plateaus has attracted much interest recently.
For weakly interacting dimer systems,
the ground state is a spin singlet liquid with only short-range correlations between spins,
and the triplet excitations require a finite excitation energy.
An excited gas of triplet magnons has two characteristic energies.
\cite{Rice}
Exchange interactions cause the transfer of excited triplets
between neighboring dimers leading to a form of kinetic energy.
They also lead to a short range repulsion between triplets
in addition to the hard core repulsion forbidding double occupancy of a dimer.
The triplet excitations are split in an external magnetic field,
and the energy of the lowest component can be driven through zero.
When the kinetic energy is dominant, which is realized in \Tl~and \K,
we have no plateaus in the magnetization curve.
\cite{Shiramura-1997}
In these case,
antiferromagnetic order perpendicular to the field appears simultaneously
with magnetization parallel to the field,
creating {\it field-induced magnetic order},
\cite{Tanaka-2001}
which can be interpreted as a condensation of the lowest lying magnon mode.
\cite{Kato,Oosawa-2002,Cavadini-1999,Cavadini-2001,Cavadini-2002,Rueegg-2002,Rueegg-2003,Nikuni,Matsumoto}
In contrast to these case,
when the interaction energy dominates,
magnetization plateaus can appear
accompanied by a magnetic superlattice.
\cite{Oshikawa}
Such plateaus have been observed
at a rational fraction of the saturation moment.
\cite{Shiramura-1998,Narumi}
In SrCu$_2$(BO$_3$)$_2$,
the frustrated form of the dimer lattice
leads to a narrow bandwidth for triplet excitations
and magnetization plateaus associated with a magnetic superlattice of localized triplets.
\cite{Miyahara,Kodama}

An unexpected feature of \N~is the absence of a magnetization plateau
at a value 1/2 of the saturation magnetization
although this would correspond to the simplest superlattice.
In this letter, an explanation of the appearance of only the values 1/4 and 3/4 is proposed.
Recently, Oosawa {\it et~al}. performed an inelastic neutron scattering experiment
at zero field in ND$_4$CuCl$_3$,
and found two almost non-dispersive excitation branches at 1.8 meV and 3 meV.
\cite{Oosawa-2003}
The feature of these excitation gaps is also reported
in specific heat measurements above the transition temperature.
\cite{Kurniawan-2}
Since the system is already ordered at zero field,
we can expect another low lying gapless branch which is hard to resolve.
Very recently, Shimaoka {\it et al.} found by using NMR
that 1/4 of the dimers are in a triplet configuration below 8.5K
in both the ordered phase and the 1/4 plateau magnetic field regions ($H<6$T),
\cite{Shimaoka}
indicating that the symmetry is lowered already above the magnetic transition temperature,
and that three weakly interacting dimer sublattices
are preformed already above the transition temperature.
These results are completely different from the SrCu$_2$(BO$_3$)$_2$ case
where a phase transition breaking translational symmetry takes place
in 1/8 plateau region.
Motivated by these results,
we propose a model composed of three distinct dimer sublattices
to account for the magnetization plateaus of \N.
In addition, we will discuss the consequences of increased interladder interactions,
e.g. due to application of an external pressure,
which can lead to the suppression of the plateaus.

%%%%%%%%%%%%%%%%%%%%%%%%%%%%%%%%%%%%%%%%%%%%%%%%%%%%%%%%%%%%%%%%%%%%%%%%%%%%%%%%
%\section{Formulation}
%%%%%%%%%%%%%%%%%%%%%%%%%%%%%%%%%%%%%%%%%%%%%%%%%%%%%%%%%%%%%%%%%%%%%%%%%%%%%%%%

%%%%%%%%%%%%%%%%%%%%%%%%%%%%%%%%%%%%%%%%%%%%%%%%%%%%%%%%%%%%%%%%%%%%%%%%%%%%%%%%
\begin{figure}[t]
\begin{center}
\includegraphics[width=7cm]{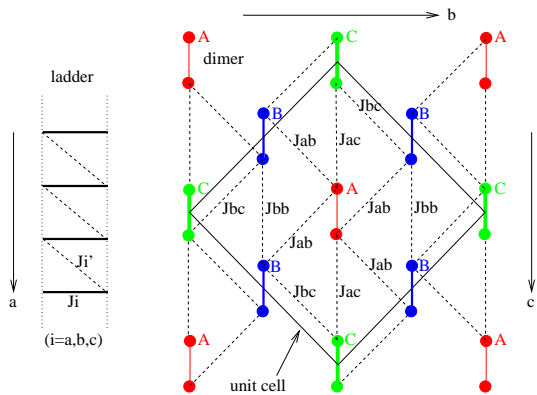}
\end{center}
\caption{
Hamiltonian of our model consisting of three distinct dimers, A, B, and C.
Circles represent $S=1/2$ spin.
Intradimer interactions are expressed by solid lines
whose width represents the strength of the interaction.
}
\label{fig:Hamiltonian}
\end{figure}
%%%%%%%%%%%%%%%%%%%%%%%%%%%%%%%%%%%%%%%%%%%%%%%%%%%%%%%%%%%%%%%%%%%%%%%%%%%%%%%%

The NH$_4^+$ ion is not spherical and is much larger than Tl$^+$ and K$^+$ ions.
The NH$_4$ molecules rotate at high temperatures,
and they freeze at low temperatures.
\cite{Rueegg-private}
In fact, elastic constant shows anomaly at 70K,
\cite{Schmidt}
and the NMR line splits below this temperature.
\cite{Shimaoka}
We speculate that the inclusion of the NH$_4^+$ ions between the ladders
gives rise to a lattice distortion at low temperatures,
modulating the exchange couplings.
It is possible, because the angle of Cu-Cl-Cu pathway for the intradimer interaction
is close to 90$^\circ$ (close to ferromagnetic exchange interaction),
and a slight change of the angle may give drastic change in the intradimer interaction.
There are various ways to distribute the three distinct dimer sublattices.
We show a possible simple model in Fig. \ref{fig:Hamiltonian}
with three distinct ladders consisting of the A, B, and C dimers.
The left and right side of each dimer is equivalent in this model.
Two-leg ladders run along the $a$ axis.
$J_i$ $(i=a,b,c)$ is an intradimer interaction
for A, B, and C dimers, respectively.
We assume $J_a \sim  0.3$ meV, $J_b\sim 1.8$ meV, and $J_c\sim 3$ meV
to reproduce the magnon excitation modes.
$J_i'$ $(i=a,b,c)$ is an interdimer interaction along the ladder.
$J_{ij}$ $(i,j=a,b,c)$ is an interdimer interaction
between $i$ and $j$ ladders.
These path ways between $S=1/2$ spins are extracted
from the crystal structure of \N.
In our model, we neglect other interactions,
because they are expected to be small.
The spin structure in the ordered phases of \N~has not yet been determined,
so we assume that its pattern is similar
to field-induced staggered order of \Tl~and \K.
Due to the modulation of the exchange couplings,
the unit cell is larger than that of \Tl~and \K,
with A and C dimers and two B dimers in the unit cell.
In the limit A, B, and C dimers are identical,
the present model reduces to the models for \Tl~and \K~in our previous work.
\cite{Matsumoto}

Taking the $z$ axis parallel to the external magnetic field,
we introduce the following variational wave function at the $i$th dimer:
\cite{Oosawa-2002-2}
\begin{equation}
\psi_i = c_{s i} |S\rangle
       + c_{\uparrow i} e^{-i\chi_i} |\uparrow\uparrow\rangle
       + c_{\downarrow i} e^{i\chi_i} |\downarrow\downarrow\rangle.
\label{eqn:Hamiltonian}
\end{equation}
Here, $|S\rangle$ is the singlet wave function.
$c_{s i}$, $c_{\uparrow i}$, and $c_{\downarrow i}$ are coefficients
expressed as
$c_{s i}=\cos{\theta_i}$,
$c_{\uparrow i}=\sin{\theta_i}\cos{\phi_i}$, and
$c_{\downarrow i}=-\sin{\theta_i}\sin{\phi_i}$.
Since the left and right side of the dimer is equivalent in our model,
the $|\uparrow\downarrow\rangle+|\uparrow\downarrow\rangle$ triplet component
does not appear in the wavefunction.
The expectation value of the spin operator of each site of a dimer is given by
\begin{eqnarray}
  \langle S_x \rangle_r &=& -\langle S_x \rangle_l =
      \frac{1}{2\sqrt{2}}\sin{2\theta_i}(\cos{\phi_i}+\sin{\phi_i})\cos{\chi_i}, \cr
  \langle S_y \rangle_r &=& - \langle S_y \rangle_l =
      \frac{1}{2\sqrt{2}}\sin{2\theta_i}(\cos{\phi_i}+\sin{\phi_i})\sin{\chi_i}, \cr
  \langle S_z \rangle_r &=& \langle S_z \rangle_l =
      \frac{1}{2}\sin^2{\theta_i}\cos{2\phi_i}.
\end{eqnarray}
Here, $\langle\cdots\rangle_{r(l)}$ represents
the expectation value on the right (left) side of the dimer.
The perpendicular ($x$ and $y$) component is staggered
(i.e. spins are aligned oppositely on $l$ and $r$ sites).
The parameter $\chi_i$ governs the rotation around the $z$-axis,
and we can determine only the relative phase, $\chi_a=\chi_c=\chi_b+\pi$,
such that the spin configuration gains the AF interladder interaction energy.
The angles, $\theta_i$ and $\phi_i$ $(i=a,b,c)$, are variational parameters
to be determined minimizing the expectation value of the Hamiltonian.
This variational method is identical to that used in our previous paper
where we introduced unitary transformations and minimized the $c$ number term
of the transformed Hamiltonian.
\cite{Matsumoto}

%%%%%%%%%%%%%%%%%%%%%%%%%%%%%%%%%%%%%%%%%%%%%%%%%%%%%%%%%%%%%%%%%%%%%%%%%%%%%%%%
%\section{Result}
%%%%%%%%%%%%%%%%%%%%%%%%%%%%%%%%%%%%%%%%%%%%%%%%%%%%%%%%%%%%%%%%%%%%%%%%%%%%%%%%

%%%%%%%%%%%%%%%%%%%%%%%%%%%%%%%%%%%%%%%%%%%%%%%%%%%%%%%%%%%%%%%%%%%%%%%%%%%%%%%%
\begin{figure}[t]
\begin{center}
\includegraphics[width=6.5cm]{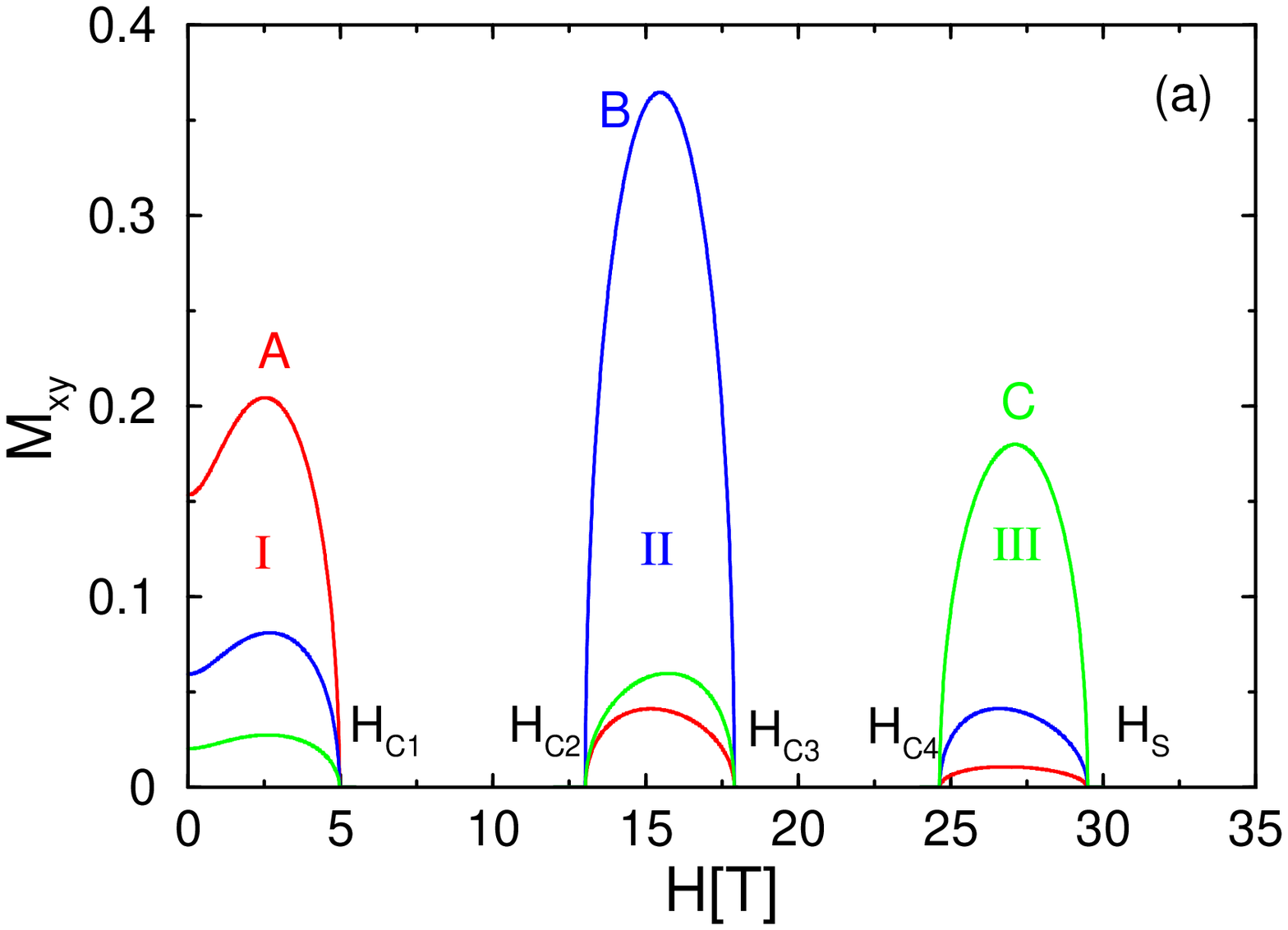}
\includegraphics[width=6.5cm]{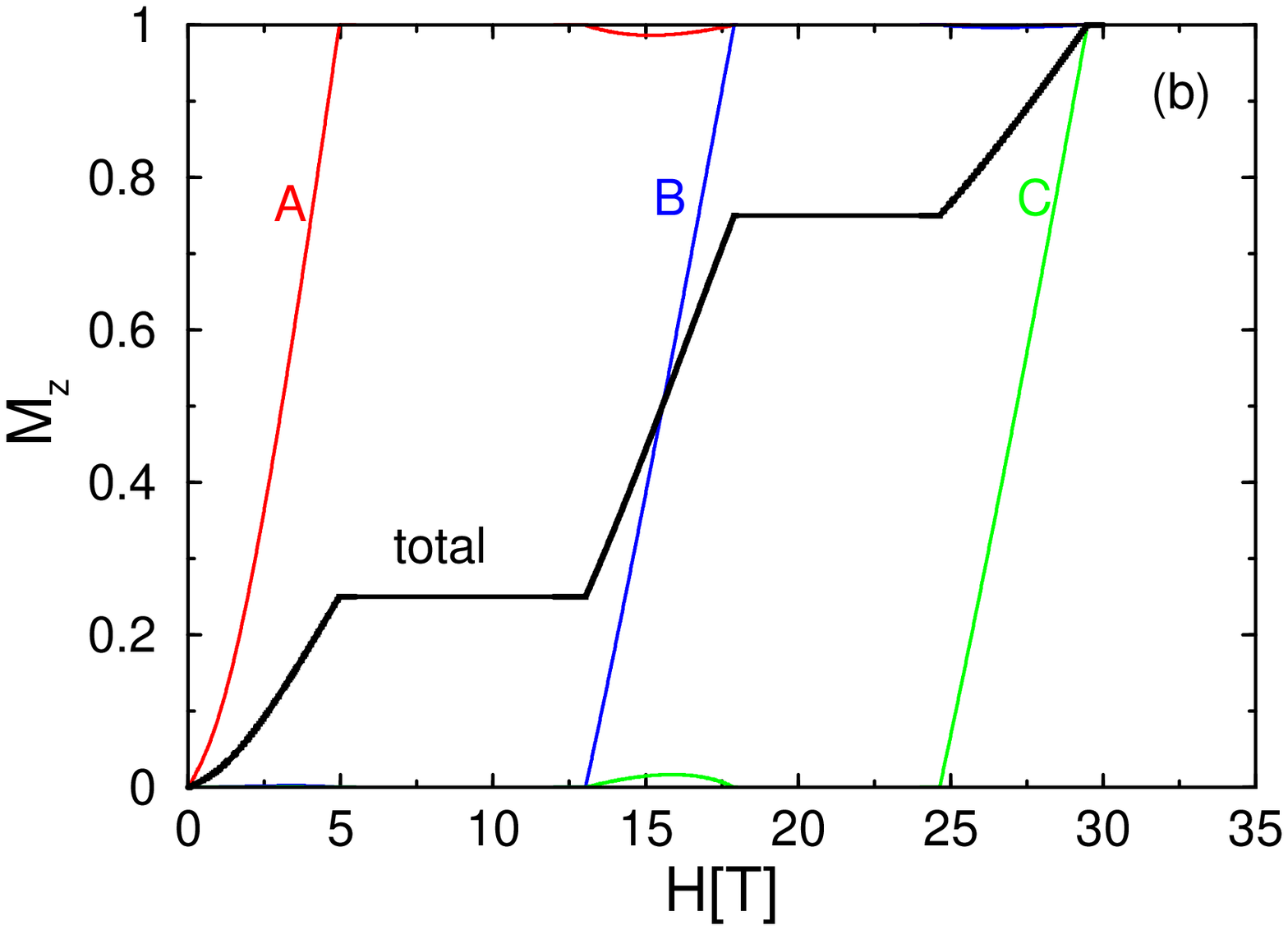}
\end{center}
\caption{
Magnetization curves at $T=0$K.
(a) Normalized staggered moment, $M_{xy}$, per volume for A, B, and C dimers.
Since there are two B dimers in the unit cell,
the staggered magnetization for the B dimer in the phase II
is about twice as large as for the A and C dimers in the phase I and III,
respectively.
(b) Uniform magnetization, $M_z$, at the A, B, and C dimer sites,
not normalized by the number of dimers.
The total magnetization is $A/4$+$B/2$+$C/4$,
and it increases by 1/4, 1/2, and 1/4
upon passing through phases I, II, and III, respectively.
Parameters were chosen to reproduce the experimental magnetization curves
in unit of meV:
$J_{a}=0.3$, $J_{b}=1.75$, $J_{c}=2.95$,
$J_{a}'=0.25$, $J_{b}'=0.2$, $J_{c}'=0.35$,
$J_{ac}=0.25$, $J_{bb}=0.1$,
$J_{ab}=0.25$, $J_{bc}=0.25$.
}
\label{fig:p=1}
\end{figure}
%%%%%%%%%%%%%%%%%%%%%%%%%%%%%%%%%%%%%%%%%%%%%%%%%%%%%%%%%%%%%%%%%%%%%%%%%%%%%%%%

Figure \ref{fig:p=1} shows the magnetization curves
obtained by the variational wavefunction.
There are three ordered phases, I, II, and III,
which are driven by the A, B, and C dimers, respectively.
At the A site,
there is a finite staggered moment ($M_{xy}$) already at zero field,
which induces staggered moment at B and C sites.
As the field increases,
this staggered moment first develops and then decreases.
At $H_{c1}$, which is a saturation field for the A dimer,
the staggered moments disappear and the A dimer is fully polarized by the field.
Since there is one A dimer in the unit cell as in Fig. \ref{fig:Hamiltonian},
we have a magnetization plateau at 1/4 for $H>H_{c1}$.
Above $H_{c2}$, which is a critical field for the B dimer,
a staggered moment develops at the B site,
inducing staggered moments at A and C sites.
Although this field region is above the saturation field for the A dimer,
the interaction between the A and B ladders
induces a staggered moment at the A site.
Accordingly, the uniform magnetization ($M_z$) at the A site
decreases somewhat in the phase II as we can see in Fig. \ref{fig:p=1}(b).
Above $H_{c3}$, which is a saturation field for the B dimer,
both B and A dimers are fully polarized.
Since there are two B dimers in the unit cell,
we have a magnetization plateau at 3/4 for $H>H_{c3}$.
Above $H_{c4}$ (critical field for the C dimer),
a staggered moment develops at the C site,
inducing the staggered moment at A and B sites.
Above $H_s$ (saturation field for the C dimer),
the all dimers are fully polarized.
Thus, the magnetization plateaus at 1/4 and 3/4 of the saturation moment
can be understood as successive quantum phase transitions driven by magnetic field,
which occur in interacting three distinct dimer sublattices.

Specific heat and high-field magnetization measurements
found three magnetically ordered phases
on the $H-T$ phase diagram.
\cite{Kurniawan-2}
It is interesting to compare it with Fig. \ref{fig:p=1}(a).
Since we expect a larger staggered magnetic moment
leads to a higher transition temperature,
we can account for the resemblance between the $H-T$ phase diagram and $M_{xy}(H)$ curves.

%%%%%%%%%%%%%%%%%%%%%%%%%%%%%%%%%%%%%%%%%%%%%%%%%%%%%%%%%%%%%%%%%%%%%%%%%%%%%%%%
\begin{figure}[t]
\begin{center}
\includegraphics[width=6.5cm]{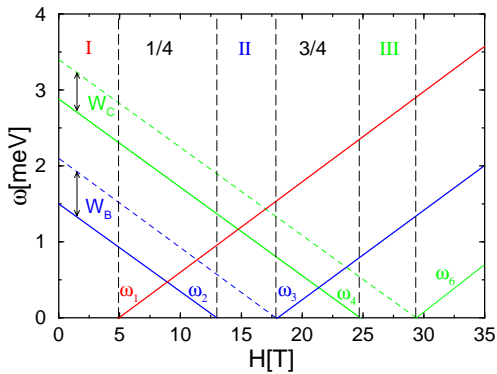}
\end{center}
\caption{
Schematic picture to understand the field dependence of excitation modes
detected by ESR measurements.
\cite{Kurniawan-1}
$\omega_1$, $\omega_3$, and $\omega_6$ are excitation modes from triplet to singlet,
while $\omega_2$ and $\omega_4$ are excitation modes from singlet to triplet.
$W_{\rm B}$ and $W_{\rm C}$ represent
bandwidth of triplet magnon excitations on the B and C dimer sublattices, respectively,
which are related to the width of the ordered phases II and III.
}
\label{fig:esr}
\end{figure}
%%%%%%%%%%%%%%%%%%%%%%%%%%%%%%%%%%%%%%%%%%%%%%%%%%%%%%%%%%%%%%%%%%%%%%%%%%%%%%%%

In our model,
the excitation branches at 1.8 meV and 3 meV
found by neutron scattering measurements are triplet excitations.
Therefore, each of them should split into three branches
when we apply external magnetic field.
It can explain the origins of the four low lying excitation modes,
$\omega_i~(i=1\sim 4)$ (see Fig. \ref{fig:esr}), detected by ESR measurements.
\cite{Kurniawan-1}
In the 1/4-plateau region,
a $\omega_1$-mode increases linearly with the field,
while the $\omega_2$-mode decreases and goes soft at $H_{c2}$.
In this field region,
each A dimer is fully polarized by the field,
and is in the lowest lying triplet $|\uparrow\uparrow\rangle$,
while the configuration of B dimer is dominated by the singlet component.
Therefore, the $\omega_1$-mode can be identified with an excitation
from a triplet to singlet on the A dimer,
and the $\omega_2$-mode with an excitation
from the singlet to the lowest lying triplet on the B dimer.
Thus, an excitation energy gap opens up in the 1/4-plateau region.
In the 3/4-plateau region,
the energy of the $\omega_3$-mode increases with the field,
while $\omega_4$-mode decreases and goes soft at $H_{c4}$.
Similarly to the 1/4-plateau region,
the $\omega_3$-mode can be identified with an excitation
from the lowest triplet to the singlet on the B dimer,
and $\omega_4$ with an excitation
from singlet to the lowest triplet on the C dimer.
$\omega_6$ in Fig. \ref{fig:esr} is an excitation
from the lowest triplet to the singlet on the C dimer,
which is not detected by ESR measurements.
There is no singlet-triplet excitation gap on the A dimer,
since an AF order appears on the A dimer already at zero field.

%%%%%%%%%%%%%%%%%%%%%%%%%%%%%%%%%%%%%%%%%%%%%%%%%%%%%%%%%%%%%%%%%%%%%%%%%%%%%%%%
\begin{figure}[t]
\begin{center}
\includegraphics[width=6.5cm]{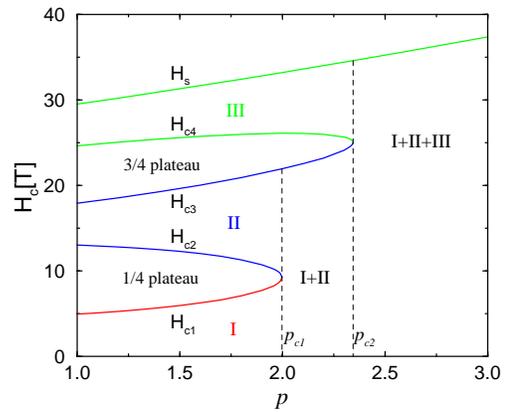}
\end{center}
\caption{
Pressure dependence of critical fields and saturation field.
Interladder interactions ($J_{ab}$, $J_{ac}$, $J_{bb}$, $J_{bc}$)
are increased by a factor of $p$
which increases linearly with pressure: $p=\alpha(P-P_0)+1$.
Here, $P$ is pressure, $P_0$ is atmosphere pressure, and $\alpha$ is a constant.
$p_{c1}=2.0$ ($p_{c2}=2.3$) corresponds to a critical pressure
on which the 1/4 (3/4) plateau vanishes.
}
\label{fig:hc}
\end{figure}
%%%%%%%%%%%%%%%%%%%%%%%%%%%%%%%%%%%%%%%%%%%%%%%%%%%%%%%%%%%%%%%%%%%%%%%%%%%%%%%%

\begin{figure}[t]
\begin{center}
\includegraphics[width=3.3cm]{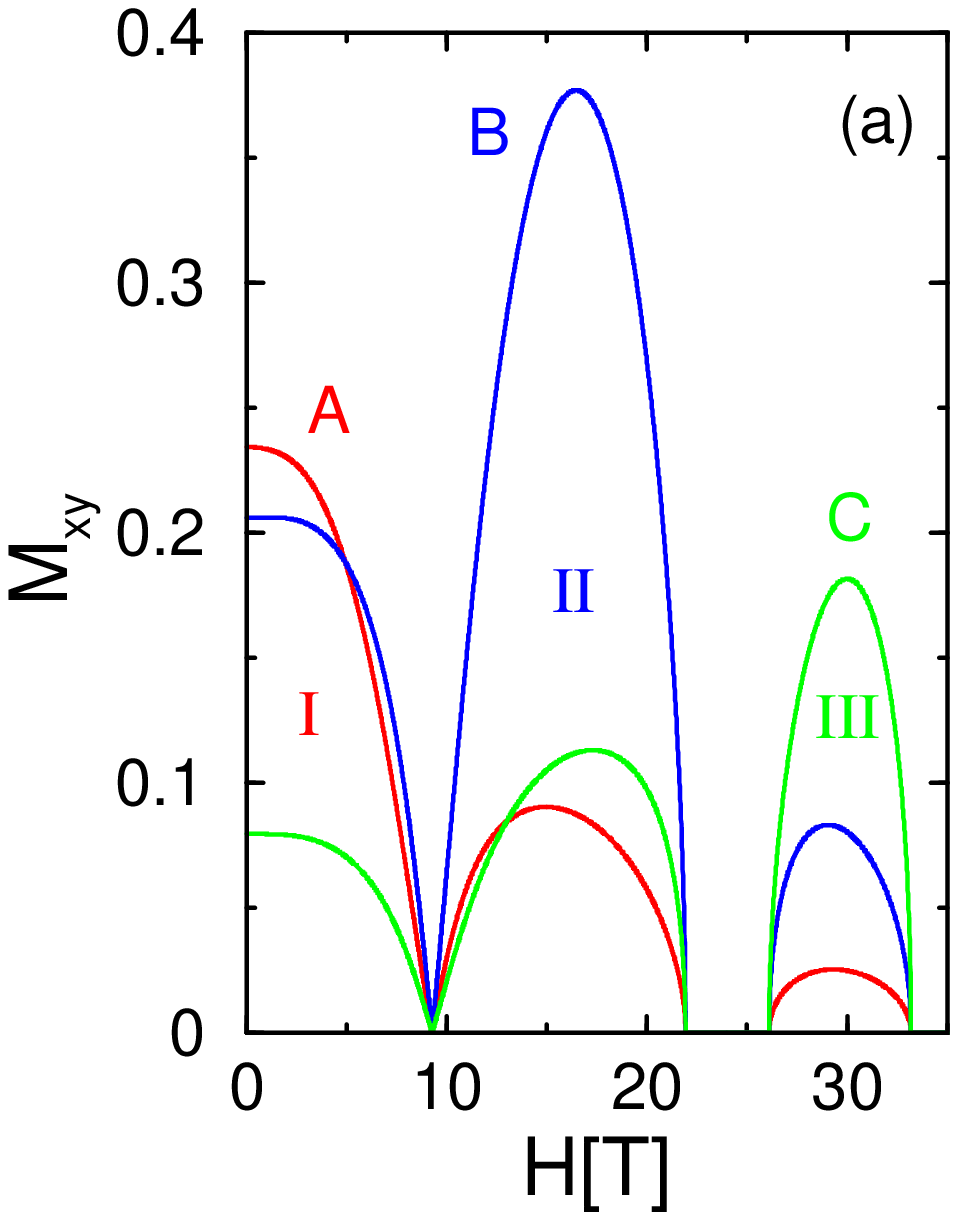}
\includegraphics[width=3.3cm]{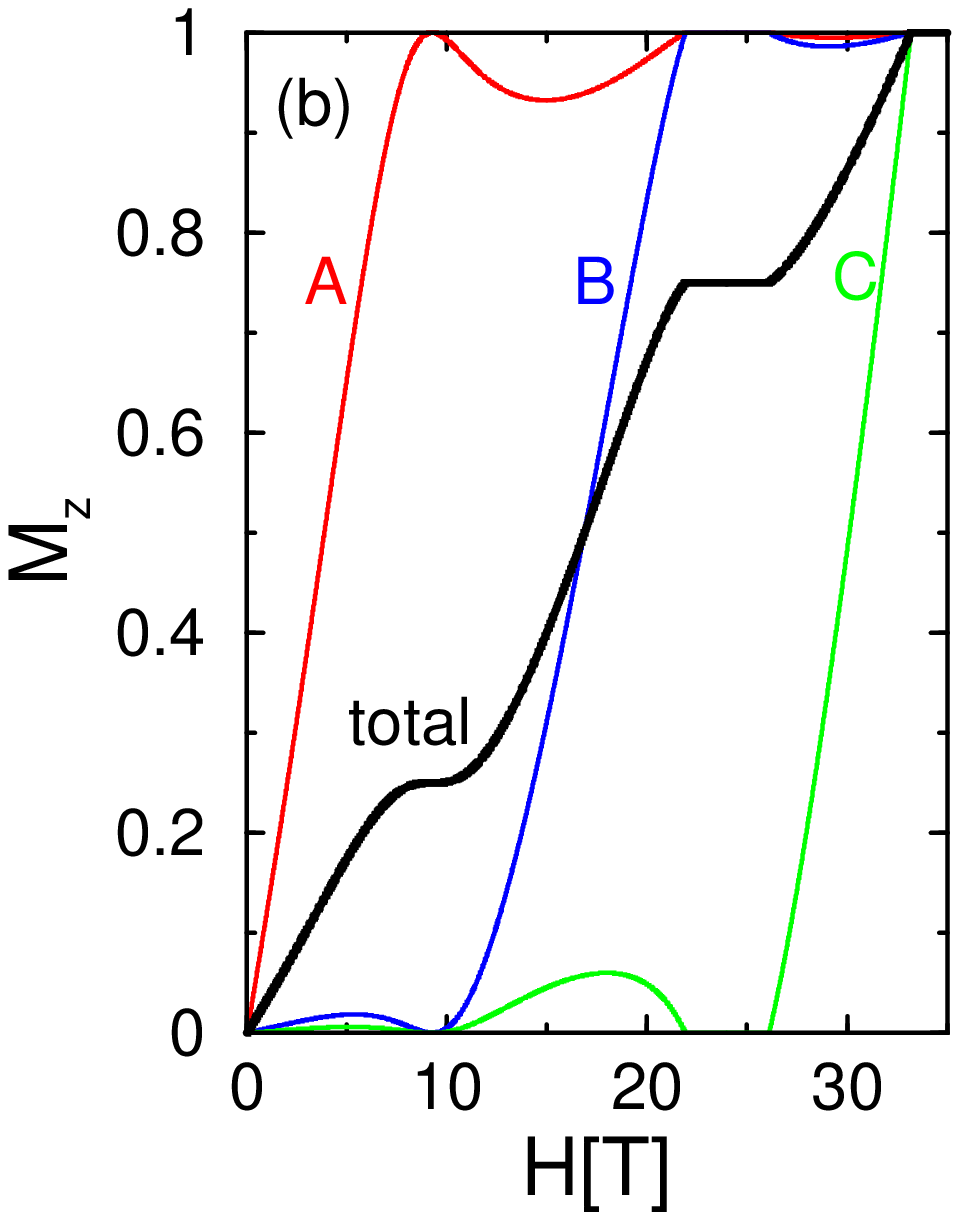}
\end{center}
\caption{
Magnetization curves under pressure $p=p_{c1}$.
(a) Staggered moment,
(b) uniform magnetization.
Interladder interactions
($J_{ac}$, $J_{bb}$, $J_{ab}$, $J_{bc}$)
are simply increased by a factor of $p_{c1}$.
The other couplings are the same as in Fig. \ref{fig:p=1}.
}
\label{fig:p=pc1}
\end{figure}

Consider now what we can predict from our model.
The critical fields ($H_{c1}$, $H_{c2}$, $H_{c3}$, $H_{c4}$)
are functions of interdimer interactions
which we can control by pressure.
If we assume that interladder interactions increase linearly with pressure,
while the intraladder interactions are constant,
we obtain the pressure dependence of the critical fields shown in Fig. \ref{fig:hc}.
As pressure increases,
the three distinct dimer sublattices couple more strongly,
and a stronger field is required to saturate spins.
Thus the plateau onset fields, $H_{c1}$, $H_{c3}$, and $H_s$,
increase with pressure,
since they correspond to saturation fields for A, B, and C dimers, respectively.
The plateau end fields, $H_{c2}$ and $H_{c4}$,
are critical fields where the lowest triplet component
on the B and C dimers are driven to zero energy.
Consider first the case of the B-sublattice.
The increase with pressure of the intrasublattice, $J_{bb}$,
broadens the triplet magnon bandwidth for the B dimer and lowers $H_{c2}$.
The coupling through the C-sublattice $J_{bc}$ acts similarly
but the coupling to the already polarized A-sublattice
acts oppositely to increase $H_{c2}$,
since the spins on the A sites are oriented parallel to the field.
The net result however is a decrease in $H_{c2}(p)$ as in Fig. \ref{fig:hc}.
This decrease leads to the disappearance of the plateau
between $H_{c1}$ and $H_{c2}$ at a critical pressure.
In Fig. \ref{fig:p=pc1}, we show the magnetizations $M_{xy}(H)$ and $M_z(H)$
at this quantum critical point.
The case of the 3/4-plateau is slightly different,
since $H_{c4}(p)$ increases with the pressure.
This occurs because there is negligible direct intrasublattice coupling
for C dimers and both the A and B sublattices are now polarized parallel to $H$.
None the less the 3/4-plateau width also narrows with increasing $p$
and eventually disappears at a higher critical pressure
above which all the ordered phases I, II, and III merge into a single phase.

%%%%%%%%%%%%%%%%%%%%%%%%%%%%%%%%%%%%%%%%%%%%%%%%%%%%%%%%%%%%%%%%%%%%%%%%%%%%%%%%
%%%%%%%%%%%%%%%%%%%%%%%%%%%%%%%%%%%%%%%%%%%%%%%%%%%%%%%%%%%%%%%%%%%%%%%%%%%%%%%%

%%%%%%%%%%%%%%%%%%%%%%%%%%%%%%%%%%%%%%%%%%%%%%%%%%%%%%%%%%%%%%%%%%%%%%%%%%%%%%%%
%\section{Conclusion}
%%%%%%%%%%%%%%%%%%%%%%%%%%%%%%%%%%%%%%%%%%%%%%%%%%%%%%%%%%%%%%%%%%%%%%%%%%%%%%%%
In summary, we have proposed a model
consisting of three distinct dimer sublattices
in order to account for the magnetization plateaus that appear in \N.
There are various way to distribute the three distinct dimer sublattices.
In this paper, we proposed a simple model
in which the dimer sublattices distribute on the $b-c$ plane
as illustrated in Fig. \ref{fig:Hamiltonian}.
In this case, triplet magnon excitations
propagate predominantly along the ladder direction.
Below $H_{c2}$, the spin structure of our model is characterized
by a wave vector (0,$\pi$,$\pi$) or (0,0,2$\pi$).
An alternate possibility is to distribute the three distinct dimer sublattices
along the ladders ($a$ axis),
which is proposed by studying NMR.
\cite{Shimaoka}
Our theory is applicable also to this case.
The distribution pattern of the three distinct dimer sublattices
is not essential to our main results.
The important assumption of our model is that
there are three weakly interacting distinct dimer sublattices
whose intradimer interactions are characterized by the values 0.3 meV, 1.8 meV, and 3 meV
observed by the neutron experiments.
\cite{Oosawa-2003}
The volume fraction of the dimers should be 1/4, 1/2, and 1/4, respectively.
Note this is consistent with the fact that
the intensity of the magnon branch at 1.8 meV
is about twice as large as the branch at 3 meV,
\cite{Oosawa-2003}
and it explains why the magnetization plateaus take place only at 1/4 and 3/4.
We made a prediction that the magnetization plateaus disappear by applying pressure,
which is supported by a recent magnetization measurement under high pressure.
\cite{Tanaka-2003}

%%%%%%%%%%%%%%%%%%%%%%%%%%%%%%%%%%%%%%%%%%%%%%%%%%%%%%%%%%%%%%%%%%%%%%%%%%%%%%%% 
%\acknowledgements
%%%%%%%%%%%%%%%%%%%%%%%%%%%%%%%%%%%%%%%%%%%%%%%%%%%%%%%%%%%%%%%%%%%%%%%%%%%%%%%%
The author expresses his sincere thanks to T. M. Rice
for valuable discussions and critical reading of the manuscript.
He is very grateful to F. Mila, C. R\"{u}egg, and H. Tanaka
for fruitful discussions.
He also would like to thank H. Kusunose, B. L\"{u}thi, B. Normand, A. Oosawa,
Y. Shimaoka, and M. Sigrist for many useful discussions.
This work is supported by Japan Society for the Promotion of Science (JSPS)
and the MaNEP program of the Swiss National Fund.

\vspace*{-6mm}

%%%%%%%%%%%%%%%%%%%%%%%%%%%%%%%%%%%%%%%%%%%%%%%%%%%%%%%%%%%%%%%%%%%%%%%%%%%%%%%%

%%%%%%%%%%%%%%%%%%%%%%%%%%%%%%%%%%%%%%%%%%%%%%%%%%%%%%%%%%%%%%%%%%%%%%%%%%%%%%%%

\end{document}